\documentclass[letterpaper, 10 pt, conference]{ieeeconf}  

\IEEEoverridecommandlockouts                              

\usepackage{color}
\usepackage{graphics} 
\usepackage{epsfig} 
\usepackage{subfigure}
\usepackage{times} 
\usepackage{amsmath} 
\usepackage{amssymb}  

\usepackage{graphicx}
\usepackage{xcolor}

\usepackage{epstopdf}

\newcommand{\R}{\mathbb{R}}

\newtheorem{definition}{Definition}[section]

\newtheorem{remarkth}[definition]{Remark}
\newenvironment{remark}{\begin{remarkth}\upshape}{\end{remarkth}}

\newcommand{\Flder}{\rightarrow}

\newcommand{\ra}{\rightarrow}

\newcommand{\lcf}{\lbrack\! \lbrack}
\newcommand{\rcf}{\rbrack\! \rbrack}

\usepackage{amssymb}

\newcommand{\proa}{A^*G \mbox{$\;$}_{\tau^*} \kern-3pt\times_\alpha
G \mbox{$\;$}_\beta \kern-3pt\times_{\tau^*} A^*G}

\def\lcf{\lbrack\! \lbrack}
\def\rcf{\rbrack\! \rbrack}

\title{\LARGE \bf
Optimal Trajectory Tracking of Nonholonomic Mechanical Systems: a geometric approach
}

\author{Aradhana Nayak$^{1}$, Rodrigo Sato Mart\'in de Almagro$^{2}$, Leonardo Colombo$^{2}$ and David Mart\'in de Diego$^{2}$
\thanks{$^{1}$A. Nayak is with Systems and Control Engineering, IIT Bombay, Mumbai India 400076 {\tt\small aradhana@sc.iitb.ac.in}}%
\thanks{$^{2}$L. Colombo, D. Mart\'in de Diego and R. Sato de Almagro are with Instituto de Ciencias Matem\'aticas (ICMAT-CSIC), Calle Nicol\'as Cabrera 15, Campus UAM, Cantoblanco, 28049, Madrid, Spain.
        {\tt\small leo.colombo@icmat.es, david.martin@icmat.es, rodrigo.sato@icmat.es}}
\thanks{*The work of A. Nayak has been partially supported by Alianza 4 Erasmus Fellowship. The work of L. Colombo, D. Mart\'in de Diego and R. Sato Mart\'in de Almagro has been partially supported by  MINECO (Spain) grant MTM2016-76072-P}}

\begin{document}

\maketitle
\thispagestyle{empty}
\pagestyle{empty}

\begin{abstract}
We study the tracking of a trajectory for a nonholonomic system by recasting the problem as an optimal control problem. The cost function is chosen to minimize the error in positions and velocities between the trajectory of a nonholonomic system and the desired reference trajectory evolving on the distribution which defines the nonholonomic constraints. We prepose a geometric framework since it describes the class of nonlinear systems under study in a coordinate-free framework. Necessary conditions for the existence of extrema are determined by the Pontryagin Minimum Principle. A nonholonomic fully actuated particle is used as  a benchmark example to show how the proposed method is applied.
\end{abstract}
%

\section{Introduction}

Nonholonomic optimal control problems arise in many engineering applications,
for instance systems with wheels, such as cars and bicycles, and systems with
blades or skates. There are thus multiple applications in the context of wheeled
motion, space or mobile robotics and robotic manipulation. The earliest work on control of nonholonomic systems is by R. W. Brockett in \cite{brocket}. A. M. Bloch \cite{Bl}, \cite{bl1} has examined several control theoretic issues which pertain to both holonomic and nonholonomic systems in a very general form. The seminal works about stabilization in nonholonomic control systems were done by A. M. Bloch, N. H. McClamroch, and M. Reyhanoglu in \cite{bl1}, \cite{bl2}, \cite{bl3}, \cite{blochnonh}, and more recently by A. Zuryev \cite{zh}. 

A geometrical dynamical system of mechanical type is completely determined by a Riemannian manifold $Q$, a kinetic energy, which is defined through the Riemannian metric $\mathcal{G}$ on the manifold and the potential forces encoded into a potential (conservative) function $V:Q\to\mathbb{R}$. These objects, together with a non-integrable distribution $\mathcal{D}\subset TQ$ on the tangent bundle of the configuration space determines a nonholonomic mechanical system.

Stabilization of an equilibrium point of a mechanical system on a Riemannian manifold has been a problem well studied in the literature from a geometric framework along the last decades (see \cite{Bl} and \cite{bullolewis} for a review on the topic). Further extensions of these results to the problem of tracking a smooth and bounded trajectory can be
found in \cite{bullolewis} where a proportional and derivative plus feed forward
(PD+FF) feedback control law is proposed for tracking a trajectory on a Riemannian manifold using error functions. 

For trajectory tracking, the usual approach of stabilization of error dynamics \cite{kodi}, \cite{aradhana1}, \cite{aradhana2}, \cite{sanyal}  cannot be utilized for nonholonomic systems. This is because there does not exist a $\mathcal{C}^1$ (even continuous) state feedback which can asymptotically stabilize the trajectory of a nonholonomic system about a desired equilibrium point. The closed loop trajectory violates Brockett's condition \cite{brocket2}, \cite{blochnonh} which states that any system of the form $\dot{x} = f(x,u)$ must have a neighborhood of zero in the image of the map $x \to f(x,u)$ for some $u$ in the control set.  This result appears in Theorem 4 in \cite{blochnonh}.

In this paper, we introduce a geometrical framework in nonholonomic mechanics to study tracking of trajectories for nonholonomic systems based on \cite{leothesis}, \cite{Cobook}, \cite{CoMa}. The application of modern tools from differential geometry in the fields of mechanics, control theory and numerical integration has led to significant progress in these research areas. For instance, the study in a geometrical formulation of the nonholonomic equations of motion has led to better understanding of different engineering problems such locomotion generation, controllability, motion planning, and trajectory tracking. 

Combining the ideas of geometric methods in control theory, nonholonomic systems and optimization techniques, in this paper, we study the underlying geometry of a tracking problem for nonholonomic systems by understanding it as an optimal control problems for mechanical systems subject to nonholonomic constraints.

Given a reference trajectory $\gamma_{r}(t)=(q_r(t), v_r(t))$ on $\mathcal{D}$ the problem studied in this work consists on finding an admissible curve $\gamma(t)\in\mathcal{D}$, solving a dynamical control system, with prescribed boundary conditions on $\mathcal{D}$ and minimizing a cost functional which involves the error  between the reference trajectory and the trajectory we want to find (in terms of both, positions and velocities), and the effort of the control inputs. This cost functional is accomplished with a weighted  terminal cost  (also known as Mayer term) which induces a constraint into the dynamics on $\mathcal{D}$.  The interval length for the cost functional $T$  may either be fixed, or appear as a degree of freedom in the optimization problem, or be time horizon. In this work, we restrict to the case when $T$ is fixed. 

To test the efficiency of the proposed method, we use a Runge Kutta integrator together with a shooting method in the solution of a trajectory optimization for a simple but challenging benchmark mechanical system: a fully actuated particle subject to a nonholonomic constraint into the dynamics. 

We propose a geometric derivation of the equations of motion for tracking a trajectory of a nonholonomic system as an optimal
control problem find we find necessary conditions via the Pontryagin Minimum Principle (PMP), where the optimal Hamiltonian is defined on the cotangent bundle of the constraint distribution. This approach allow for the reduction in the degrees of freedom of the equations for the optimal control problem, compared with typical methods describing the dynamics of a nonholonomic system, as the ones arising from  the application of Lagrange-d'Alembert principle. The main advantages in this geometric framework consist in the use of a basis of vector fields on $\mathcal{D}$ allowing the reduction of some degrees of freedom in the dynamics for a nonholonomic mechanical system.  
The paper is structured as follows: we introduce mechanical systems on a manifold, connections on a Riemannian manifold and the geometry of nonholonomic dynamical systems on Section \ref{sec2}, together with the example we used as benchmark the nonholonomic particle. Section \ref{section3} introduces the details of the problem under study motivated by the non-existence of a $\mathcal{C}^1$ feedback control to asymptotically stabilize the error dynamics in nonholonomic systems. Necessary conditions for the existence of extrema in the proposed optimal control problem are studied from the PMP  in Section \ref{sec4}. We also show numerical results and analyze the results we obtain. 

\section{Nonholonomic Mechanical Systems}\label{sec2}
\subsection{Preliminaries}
\label{section2} Let $Q$ be a $n$-dimensional differentiable
manifold with local coordinates $(q^i)$, with $1\leq i\leq n$, the
configuration space of a mechanical system. Denote by $TQ$ its
tangent bundle with induced local coordinates $(q^i, \dot{q}^i)$.
Given a Lagrangian function $L:TQ\rightarrow \R$, its Euler-Lagrange
equations are
\begin{equation}\label{qwer}
\frac{d}{dt}\left(\frac{\partial L}{\partial\dot
q^i}\right)-\frac{\partial L}{\partial q^i}=0, \quad 1\leq i\leq n.
\end{equation}
These equations determine a system of implicit second-order
differential equations in general. If we assume that the Lagrangian is regular,
that is, the ${n\times n}$ matrix $\left(\frac{\partial^{2} L}{\partial \dot q^i
\partial \dot q^j}\right)$ is non-degenerate, the local existence and uniqueness of solutions is guaranteed for any given initial condition.

Vector fields are used to calculate the directional derivative of a function defined on $Q$. In the realm of differential geometry a more general operator is defined to perform derivation of a wider range of geometric objects (tensors). This operator is called connection (linear, covariant, or affine connection). The definition of the connection is a wish list of properties which it is expected to have
\begin{definition} An (affine) connection on a smooth manifold $Q$ is a map which takes a pair consisting of a vector (or a vector field), and a $(p,q)$-tensor field, $T$, and returns a $(p,q)$-tensor field, such that it satisfies the following axioms
\begin{itemize}
\item $\nabla_{X}f=X(f)$, for $f\in\mathcal{C}^{\infty}(Q)$,
\item $\nabla_{X}(T+S)=\nabla_{X}T+\nabla_{X}S$, for $T$ and $S$ tensors of the same type,
\item $\nabla_{X}T(f,g)=(\nabla_XT)(f,g)+T(\nabla_Xf,g)+T(f,\nabla_Xg)$.
\end{itemize}
\end{definition}

This definition of a connection is complete, i.e., this
list of properties results in a uniquely defined geometric
operator; however, an extra structure on the manifold is
needed to define this object in a chart. To do so, we need to
know how it acts on the basis of the tangent vector space. The result is a tangent vector field, and at each point it is spanned by the basis of the tangent space at that point
$\displaystyle{\nabla_{\frac{\partial}{\partial q^{i}}}\left(\frac{\partial}{\partial q^{j}}\right)=\Gamma^{k}_{ij}\frac{\partial}{\partial q^{k}}.}$

Denote by $\mathfrak{X}(Q)$ the set of vector fields on $Q$. A metric $\mathcal{G}$ on a smooth manifold is a $(0,2)$-tensor field satisfying 
\begin{itemize}\item Symmetry: $\mathcal{G}(X,Y)=\mathcal{G}(Y,X)$ $X,Y\in\mathfrak{X}(Q)$,
\item Non-degeneracy: $\mathcal{G}(X,Y)=0$ if and only if when $X=0$ then $Y=0$.
\end{itemize}

Locally, the metric is
determined by the matrix $M=(\mathcal{G}_{ij})_{1\leq i, j\leq n}$
where $\mathcal{G}_{ij}=\mathcal{G}(\partial/\partial q^i,
\partial/\partial q^j)$.

Using the metric $\mathcal{G}$ we may compute the Christoffel symbols associated with the metric as $$\Gamma_{ij}^{k}=\left(\mathcal{G}^{-1}\right)_{ks}\left(\frac{\partial\mathcal{G}_{sj}}{\partial q^{i}}+\frac{\partial\mathcal{G}_{si}}{\partial q^j}+\frac{\mathcal{G}_ij}{\partial q^{s}}\right)$$ where $\mathcal{G}^{-1}$ is defined as the inverse of the metric with components determined by the inverse matrix of $M$.

\subsection{Nonholonomic mechanical systems}
Most nonholonomic systems have linear constraints, and these are the ones we will consider. Linear constraints on the
velocities (or Pfaffian constraints) are locally given by equations of the form $\phi^{a}(q^i, \dot{q}^i)=\mu^a_i(q)\dot{q}^i=0, \quad 1\leq a\leq
m$, depending, in general, on their configuration coordinates and  their
velocities. {}From an intrinsic point of view, the linear
constraints are defined by a regular distribution ${\mathcal D}$ on
$Q$ of constant rank $n-m$ such that the annihilator of ${\mathcal
D}$ is locally given at each point of $Q$ by
${\mathcal D}^o_{q} = \operatorname{span}\left\{ \mu^{a}(q)=\mu_i^{a}dq^i \; ; 1 \leq a
\leq m \right\}$ where the one-forms $\mu^{a}$ are independent at each point of $Q$.

Now we restrict ourselves to the case of nonholonomic mechanical systems where the Lagrangian is of mechanical type, that is, a Lagrangian systems $L:TQ\to\R$ defined by \[
L(v_q)=\frac{1}{2}\mathcal{G}(v_q, v_q) - V(q),
\]
with $v_q\in T_qQ$, where $\mathcal{G}$ denotes a Riemannian metric on the configuration
space $Q$ representing the kinetic energy of the systems and
$V:Q\ra\R$ is a potential function. 

Next, assume that the system is subject to nonholonomic constraints, defined by a regular distribution $\mathcal{D}$ on $Q$ with corank$(\mathcal{D})=m$.
Denote by $\tau_{\mathcal{D}}:\mathcal{D}\ra Q$ the canonical
projection of $\mathcal{D}$ onto $Q$ and by 
$\Gamma(\tau_{\mathcal{D}})$ the set of sections of $\tau_{D}$ which
in this case is just the set of vector fields $\mathfrak{X}(Q)$
taking values on $\mathcal{D}.$ If $X, Y\in\mathfrak{X}(Q),$ then
$[X,Y]$ denotes the standard Lie bracket of vector fields.

\begin{definition}\label{nonholonomicsystem}
A \textit{nonholonomic mechanical system} on a smooth manifold $Q$ is given
by the triple $(\mathcal{G}, V, \mathcal{D})$, where $\mathcal{G}$ is
a Riemannian metric on $Q,$ representing the kinetic energy of the
system, $V:Q\ra\R$ is a smooth function representing the potential
energy and $\mathcal{D}$ a non-integrable regular distribution on $Q$
representing the nonholonomic constraints.
\end{definition}

Given $X,Y\in\Gamma(\tau_{\mathcal{D}})$ that is,
$X(x)\in\mathcal{D}_{x}$ and $Y(x)\in\mathcal{D}_{x}$ for all $x\in
Q,$ then it may happen that $[X,Y]\notin\Gamma(\tau_{\mathcal{D}})$
since $\mathcal{D}$ is nonintegrable. We want to obtain a bracket definition for sections of $\mathcal{D}.$
Using the Riemannian metric $\mathcal{G}$ we can define two
complementary orthogonal projectors ${\mathcal P}\colon TQ\to {\mathcal D}$ and ${\mathcal Q}\colon TQ\to {\mathcal
D}^{\perp},$ with respect to the tangent bundle orthogonal decomposition $\mathcal{D}\oplus\mathcal{D}^{\perp}=TQ$. Therefore, given $X,Y\in\Gamma(\tau_{\mathcal{D}})$ we define the
\textit{nonholonomic bracket}
$\lcf\cdot,\cdot\rcf:\Gamma(\tau_{\mathcal{D}})\times\Gamma(\tau_{\mathcal{D}})\rightarrow\Gamma(\tau_{\mathcal{D}})$
as $\lcf X_A,X_B\rcf:=\mathcal{P}[X_A,X_B]$. This Lie bracket verifies the usual properties of a Lie bracket except the Jacobi identity (see \cite{BlCoGuMdD}, \cite{leo2} for example). 

\begin{definition}
Consider the restriction of the Riemannian metric $\mathcal{G}$ to
the distribution $\mathcal{D}$, $\mathcal{G}^{\mathcal{D}}:\mathcal{D}\times_{Q}\mathcal{D}\ra\R$
and define the \textit{Levi-Civita connection}
$\displaystyle{\nabla^{\mathcal{G}^{\mathcal{D}}}:\Gamma(\tau_{\mathcal{D}})\times\Gamma(\tau_{\mathcal{D}})\ra\Gamma(\tau_{\mathcal{D}})}$
determined by the following two properties:

\begin{enumerate}
\item $\lcf X,Y\rcf=\nabla_{X}^{\mathcal{G}^{\mathcal{D}}}Y-\nabla_{Y}^{\mathcal{G}^{\mathcal{D}}}X,$
\item $X(\mathcal{G}^{\mathcal{D}}(Y,Z))=\mathcal{G}^{\mathcal{D}}(\nabla_{X}^{\mathcal{G}^{\mathcal{D}}}Y,Z)+\mathcal{G}^{\mathcal{D}}(Y,\nabla_{X}^{\mathcal{G}^{\mathcal{D}}}Z).$
\end{enumerate}

\end{definition}

Let $(q^{i})$ be local coordinates on $Q$ and $\{e_{A}\}$ be independent vector fields on
$\Gamma(\tau_{D})$ (that is, $e_A(x)\in {\mathcal D}_x$) such that $$\mathcal{D}_{x}=\hbox{span }\{e_{A}(x)\},
\quad x\in U\subset Q.$$ Then, we can determine the \textit{Christoffel
symbols} $\Gamma_{BC}^{A}$ of the connection
$\nabla^{\mathcal{G}^{\mathcal{D}}}$ by $\displaystyle{\nabla_{e_{B}}^{\mathcal{G}^{\mathcal{D}}}e_{C}=\Gamma_{BC}^{A}(q)e_{A}.}$

As when we work in tangent bundles, it is possible to determine the \textit{Christoffel
symbols} associated with the connection $\nabla^{{\mathcal
G}^{\mathcal{D}}}$ by $\nabla^{{\mathcal G}^{\mathcal{D}}}_{e_B}{e_C}=\Gamma^A_{BC}e_A$. Note that the coefficients $\Gamma_{AB}^{C}$ of the connection  $\nabla^{{\mathcal
G}^{\mathcal{D}}}$ are (see \cite{maria2} for details) \begin{equation}\label{relation}\Gamma_{AB}^{C}=\frac{1}{2}(\mathcal{C}_{CA}^{B}+\mathcal{C}_{CB}^{A}+\mathcal{C}_{AB}^{C})\end{equation}where the constant structures $\mathcal{C}_{AB}^{C}$ are defined as $\lcf X_A,X_B\rcf=\mathcal{C}_{AB}^{C}X_C$.

\begin{definition}
A curve $\gamma:I\subset\R\ra\mathcal{D}$ is \textit{admissible} if
$\displaystyle{\gamma(t)=\frac{d\sigma}{dt}(t)}$, where $\tau_{\mathcal{D}}\circ\gamma=\sigma$.
\end{definition}

Given local coordinates on $Q,$ $(q^{i})$ with $i=1,\ldots,n;$ and
$\{e_{A}\}$ sections on $\Gamma(\tau_{\mathcal{D}})$, with $A=1,\ldots,n-m$, such that
$\displaystyle{e_{A}=\rho_{A}^{i}(q)\frac{\partial}{\partial
q^{i}}}$ we introduce induced coordinates $(q^{i},v^{A})$ on
$\mathcal{D}$, where, if $e\in\mathcal{D}_{x}$ then
$e=v^{A}e_{A}(x).$ Therefore, $\gamma(t)=(q^{i}(t),v^{A}(t))$ is
admissible if
$$\dot{q}^{i}(t)=\rho_{A}^{i}(q(t))v^{A}(t).$$

Consider the restricted Lagrangian function
$\ell:\mathcal{D}\rightarrow\mathbb{R},$
$$\ell(v)=\frac{1}{2}\mathcal{G}^{\mathcal{D}}(v,v)-V(\tau_{D}(v)),\hbox{ with }  v\in\mathcal{D}.$$

\begin{definition}
A \textit{solution of the nonholonomic problem} is an admissible
curve $\gamma:I\rightarrow\mathcal{D}$ such that
$$\nabla_{\gamma(t)}^{\mathcal{G}^{\mathcal{D}}}\gamma(t)+grad_{\mathcal{G}^{\mathcal{D}}}V(\tau_{\mathcal{D}}(\gamma(t)))=0.$$
\end{definition} Here the section $grad_{{\mathcal G}^{\mathcal{D}}}V\in\Gamma(\tau_{\mathcal{D}})$ is characterized by \[ {{\mathcal G}^{\mathcal{D}}}(grad_{{\mathcal
G}^{\mathcal{D}}}V, X) = X(V), \; \; \mbox{ for  every } X \in
\Gamma(\tau_{\mathcal{D}}).
\]

These equations are equivalent to the \textit{nonholonomic
equations}. Locally, these equations are given by
\begin{align}
\dot{q}^{i}&=\rho_{A}^{i}(q)v^{A}\label{eqq1}\\
\dot{v}^{C}&=-\Gamma_{AB}^{C}v^{A}v^{B}-(\mathcal{G}^{\mathcal{D}})^{CB}\rho_{B}^{i}(q)\frac{\partial V}{\partial q^{i}},\label{eqq2}
\end{align} where $(\mathcal{G}^{\mathcal{D}})^{AB}$ denotes the coefficients of the inverse matrix of $(\mathcal{G}^{\mathcal{D}})_{AB}$ where $\mathcal{G}^{\mathcal{D}}(e_{A},e_{B})=(\mathcal{G}^{\mathcal{D}})_{AB}.$

\begin{remark}
The nonholonomic  equations only depend on the coordinates
$(q^{i},v^{A})$ on $\mathcal{D}$.  Therefore the nonholonomic
equations are free of Lagrange multipliers. These equations are
equivalent to the  \textit{nonholonomic Hamel equations} (see
\cite{BloZen}, for example, and references therein).

\end{remark}

\subsection{Example: The nonholonomic particle}\label{example}

Consider a particle of unit mass evolving in $Q=\R^3$ with Lagrangian $\displaystyle{
L(x,y,z,\dot x,\dot y,\dot z)=\frac{1}{2}(\dot x^2+\dot y^2+\dot z^2)}$, and subject to the constraint $\dot x+y\,\dot z=0$.

 This nonholonomic system is defined by the annhilation of the one-form $\mu(x,y,z)=(1,0,y)$. The nonholonomic equations, derived from the Lagrange-d'Alembert principle, are given by
\begin{align}\label{eeqq2}
\dot v_x=&\lambda,\quad\quad\quad v_x=\dot x,\qquad \dot v_y=0,\quad\quad v_y=\dot y,\\
\dot v_z=&y\lambda,\quad\quad\,\,\, v_z=\dot z,\qquad v_x+y\,v_z=0, \nonumber
\end{align} which, after substituting the Lagrange multiplier $\displaystyle{\lambda=-\frac{v_zv_y}{1+y^2}}$, lead to 

\begin{align}
\dot v_x=&-\frac{y}{1+y^2}v_z v_y, \quad\dot v_z=-\frac{1}{1+y^2}\,v_z\,v_y, \quad  \label{NhPequationsSL}\\ v_z=&\dot z, \quad v_y=\dot y,\quad v_x=\dot x, \quad \dot v_y=0,\label{NhPequationsSL2}\end{align}

such that $v_x+y\,v_z=0$.  Let $\mathcal{D}\subset T\R^{3}\simeq \R^3\times\R^3$ denote the nonholonomic distribution corresponding to this system.  Then these equations define a time-continuous flow $F_t:\mathcal{D}\Flder \mathcal{D}$, i.e. $F_t((q(0),v(0)))=(q(t),v(t))$, where $q(t)=(x(t),y(t),z(t))^T$ and $v(t)=(v_x(t),v_y(t),v_z(t))^T$, $(q(0),v(0))\in\mathcal{D}$.

 The distribution $\mathcal{D}$ is determined by  $\mathcal{D} = \hbox{span} \{Y_1, Y_2\}= \hbox{span}\Big{\{} \frac{\partial}{\partial y}, \frac{\partial}{\partial z}- y \frac{\partial}{\partial x}\Big{\}} $. Then, $\mathcal{D}^{\perp}=\{\frac{\partial}{\partial x}+y\frac{\partial}{\partial z}\}$. Let $(x,y,z,v^1,v^2)$ be induced coordinates on $\mathcal{D}$.
   
Given the  vector fields $Y_1$ and $Y_2$ generating the distribution  we obtain the relations for $q\in \R^3$
\begin{eqnarray*}
Y_1(q)&=&\rho_1^1(q)\frac{\partial}{\partial x}+\rho_1^2(q)\frac{\partial}{\partial y}+\rho_1^3(q)\frac{\partial}{\partial z},\\
Y_2(q)&=&\rho_2^1(q)\frac{\partial}{\partial x}+\rho_2^2(q)\frac{\partial}{\partial y}+\rho_2^3(q)\frac{\partial}{\partial z}.
\end{eqnarray*} Then, $\rho_{1}^{1}= \rho_{1}^{3}=\rho_2^2=0, \quad\rho_{1}^{2}=\rho_2^3=1,\quad\rho_{2}^{1}=-y$.

 Each element $e\in {\mathcal D}_q$ is expressed as a linear combination of these vector fields: $e=v^1 Y_1(q)+ v^2 Y_2(q), \quad q\in \R^3$. Therefore, the vector subbundle $\tau_{\mathcal{D}}:{\mathcal D}\rightarrow \R^3$ is locally described by the coordinates $(x,y,\theta; v^1, v^2)$; the first three for the base and the last two, for the fibers. 
Observe that 
\[
e=v^1\frac{\partial}{\partial y}+ v^2\left(\frac{\partial}{\partial z}-y\frac{\partial}{\partial x}
\right)
\]
and, in consequence, ${\mathcal D}$ is described by the conditions (admissibility conditions): $
\dot{x}=-yv^2,\quad\dot{y}=v^1,\quad\dot{z}=v^2$ as  a vector subbundle of $TQ$ where $v^1$ and $v^2$ are the adapted velocities relative to the basis of $\mathcal{D}$ defined before.

The nonholonomic bracket given
by $\lcf\cdot,\cdot\rcf=\mathcal{P}([\cdot,\cdot])$ satisfies $$\lcf
Y_1,Y_2\rcf=\mathcal{P}[Y_1,Y_2]=\mathcal{P}\left(-\frac{\partial}{\partial x}\right)=\frac{y}{1+y^2}\left(\frac{\partial}{\partial z}-y\frac{\partial}{\partial x}\right).$$

   
  Therefore, by using \eqref{relation} all the Christoffel symbols for the connection $\displaystyle{\nabla^{\mathcal{G}^{\mathcal{D}}}}$ vanish except $\Gamma_{12}^{2}$ which is given by $\displaystyle{\Gamma_{12}^{2}=\frac{y}{1+y^2}}$.
  The restriction of the Lagrangian function $L$ on $\mathcal{D}$ in the adapted coordinates $(v^1,v^2)$ is given by $$\ell(x,y,z, y_1,y_2)=\frac{1}{2}\left((v^1)^2+(v^2)^2(y^2+1)\right).$$

 Then, the nonholonomic equations for the constrained particle are given by
   \begin{equation}\label{eqq3}
     \dot{v}^1= 0,\quad
     \dot{v}^2= -\frac{y}{1+y^2}v^1v^2
   \end{equation} together with  the admissibility conditions $\dot{x}= -y v^2$,
     $\dot{y}= v^1$ and 
     $\dot{z}= v^2.$
      Then these equations define a time-continuous flow $F_t^{\mathcal{D}}:\mathcal{D}\Flder \mathcal{D}$, i.e. $F_t((q(0),v(0)))=(q(t),v(t))$, where $q(t)=(x(t),y(t),z(t))^T$ and $v(t)=(v_1(t),v_2(t))^T$, $(q(0),v(0))\in\mathcal{D}$. 
      

The previous systems can be integrated explicitly, and solutions are given by: \begin{align}
x(t)=&x_0-\frac{c_2}{c_1}\sqrt{(y_0+c_1t)^{2}+1},\quad y(t)=c_1t+y_0\nonumber\\ 
z(t)=&\frac{c_2}{2c_1}\left((y_0+c_1t)\sqrt{(y_0+c_1t)^{2}+1}+\sinh^{-1}(y_0+c_1t)\right)\nonumber\\ &-z_0c_2,\label{integral}\\
v^{1}(t)=&c_1,\quad v^{2}(t)=\frac{c_2}{\sqrt{(y_0+c_1t)^{2}+1}},\nonumber 
\end{align} for $x_0,y_0,z_0,c_1,c_2$ constants to be determined by the initial conditions.

\begin{remark}\label{remarkc=0}
Note that previous equations have a singularity at $c_1=0$. The constant $c_1$ arrises from the equation for $\dot{v}^{1}$. If $c_1=0$, and therefore ${v}^{1}(t)=0$, then the solution for the system of equations is given by $x(t)=-y_0v_0^{2}t+x_0$, $y(t)=y_0$, $z(t)=v_0^{2}t+z_0$, $v^{2}(t)=v_0^{2}$, where $x_0,y_0,z_0,v_0^2$ are constants. \hfill$\diamond$
\end{remark}
\section{Optimal trajectory tracking problem}\label{section3}

The purpose of this section is to present the tracking problem for nonholonomic systems as an optimal control problem. The objective is the tracking of a suitable reference trajectory $\Upsilon(t)$ for a mechanical system with nonholonomic velocity constraints as described in the previous section. It is assumed that $\Upsilon(t)\in\mathcal{D}$. 

We will analyze the case when the dimension of the inputs set or control distribution is equal to the rank of $\mathcal{D}$. If the rank of $\mathcal{D}$ is equal to the dimension of the control distribution, the system will be called a \textit{fully actuated nonholonomic system}.

\begin{definition}
A \textit{solution of a fully actuated nonholonomic problem} is an
admissible curve $\gamma:I\rightarrow\mathcal{D}$ such that
$$\nabla_{\gamma(t)}^{\mathcal{G}^{\mathcal{D}}}\gamma(t)+grad_{\mathcal{G}^{\mathcal{D}}}V(\tau_{\mathcal{D}}(\gamma(t)))\in \Gamma (\tau_D),$$
or, equivalently, 
$$\nabla_{\gamma(t)}^{\mathcal{G}^{\mathcal{D}}}\gamma(t)+grad_{\mathcal{G}^{\mathcal{D}}}V(\tau_{\mathcal{D}}(\gamma(t)))=u^{A}(t)e_{A}(\tau_{\mathcal{D}}(\gamma(t))),$$
where $u^{A}$ are the control inputs.
\end{definition}

Locally, the above equations are given by 
\begin{eqnarray}
\dot{q}^{i}&=&\rho_{A}^{i}v^{A}\label{control1}\\
\dot{v}^{A}&=&-\Gamma_{CB}^{A}v^{C}v^{B}-(\mathcal{G}^{\mathcal{D}})^{AB}\rho_{B}^{i}(q)\frac{\partial V}{\partial q^{i}}+u^{A}.\label{control2}
\end{eqnarray}

As we mentioned in the Introduction, For trajectory tracking, the usual approach of stabilization of error dynamics \cite{kodi}, \cite{aradhana1}, \cite{aradhana2}, \cite{sanyal}  cannot be utilized for nonholonomic systems because the closed loop trajectory violates Brockett's condition. A common approach to trajectory tracking for nonholonomic systems found in the literature is the backstepping procedure \cite{Ha}, \cite{nijmeijer97}. This approach is done on a per example basis, in particular, mobile robots or unycicle models.  In \cite{Ha}, \cite{nijmeijer97} the error dynamics of the unicycle model is shown to be in strict feedback form. Thereafter, integrator backstepping is employed to choose an appropriate Lyapunov function for stabilization of the error dynamics. This error dynamics does not evolve on the constrained manifold (unlike our approach). Therefore, Brockett's condition is not violated. However, since $\rho^i_A(q)$ is unknown in a general framework, the approach can not be generalized to solve the tracking problem for a general nonholonomic system with our method and then backstepping needs to be studied for each system. 


So we propose a new approach to consider tracking problem as an optimal control problem and we call this \textit{optimal tracking}.

In the following, we shall assume that all the control systems under consideration are controllable in the configuration space, that is, for any two
points $q_0$ and $q_f$ in the configuration space $Q$, there exists
an admissible control $u(t)$ defined on the control manifold
$\mathcal{U}\subseteq\R^{n}$ such that the system with initial condition $q_0$
reaches the point $q_f$ at time $T$ (see \cite{Bl} for
more details). Given a cost function $
\mathcal{C}:\mathcal{D}\times \mathcal{U}\rightarrow\mathbb{R}$ the \textit{optimal control problem} consists of finding an admissible curve
$\gamma:I\rightarrow\mathcal{D}$ which is a solution of the fully actuated
nonholonomic problem given initial and final boundary conditions on
$\mathcal{D}$ and minimizing the cost functional
$$\mathcal{J}(\gamma(t),u(t)):=\int_{0}^{T}\mathcal{C}(\gamma(t),u(t))dt.$$

For trajectory tracking of a nonholonomic system we consider the following problem

\textbf{Problem (optimal trajectory tracking):} Given a reference trajectory $\gamma_{r}(t)=(q_r(t), v_r(t))$ on $\mathcal{D}$, find an admissible curve $\gamma(t)\in\mathcal{D}$, solving \eqref{control1}-\eqref{control2}, with prescribed boundary conditions on $\mathcal{D}$ and minimizing the cost functional 
\begin{align*}\mathcal{J}(\gamma(t)) =& \frac{1}{2}\int_{0}^{T} \left(||\gamma(t)- \gamma_{r}(t)||^2 + \epsilon ||u^A||^2 \right)\,dt+\omega\Phi(\gamma(T))\\
& =\frac{1}{2}\int_{0}^{T} \left(||q^{i}(t)- q^{i}_{r}(t)||^2 +||v^{A}(t)-v^{A}_{r}(t)||^2 \right.\\&\left.\quad+\epsilon ||u^A||^2 \right)\,dt+\omega\Phi(\gamma(T))
\end{align*} where $\epsilon>0$ is a regularization parameter, $\Phi:TQ\to\mathbb{R}$ is a terminal cost (Mayer term), $\omega>0$ is a weight for the terminal cost. $\mathcal{C}$ and $\Phi$ are assumed to be continuously differentiable functions, and the final state $\gamma(T)$ is required to fulfill a constraint $r(\gamma(T), \gamma_{r})=0$ with $r:\mathcal{D}\times\mathcal{D}\to\mathbb{R}^{d}$ and $\gamma_r\in\mathcal{D}$ given. The interval length $T$ may either be fixed, or appear as degree of freedom in the optimization problem. In this work we restrict to the case when $T$ is fixed.

\begin{remark}\label{rksingular}
Note that if $\epsilon=0$ then the optimal control problem turns into a singular optimal control problem (see \cite{OTT} Section $3.2$)\hfill$\diamond$.
\end{remark}
\section{Necessary conditions for optimality}\label{sec4}

In this section we apply Pontryagin's minimization principle to the optimal tracking problem. The Hamiltonian $\mathcal{H}:T^{*}\mathcal{D}\times \mathcal{U}\to\mathbb{R}$ for the problem is given by
\begin{align} \mathcal{H}(q,v, \lambda,\mu,u)=& \mathcal{J}(q^i,v^A,u^A) + \lambda_i \rho_{A}^{i}(q)v^{A}\label{hamiltonian}\\ 
&+ \mu_A \dot{v}^A(q^{i},v^{A},u^{A})\nonumber
\end{align} where $\dot{v}^{A}$ comes from equation \eqref{control2}. Note that $\lambda_i$ and $\mu_A$ are the costate variables or Lagrange multipliers. The last two terms in \eqref{hamiltonian} corresponds with the nonholonomic dynamics given in equations \eqref{eqq1} and \eqref{eqq2} paired with the costate variables, which represents the standard construction of the Hamiltonian for the PMP. Also note that $\mathcal{H}$ is defined on a subset of $T^{*}(TQ)$.

Denote by $t\mapsto u^{\star}(t)$ a curve that satisfies along a trajectory $t\mapsto (q(t),v(t),\lambda(t),\mu(t))\in T^{*}\mathcal{D}$, $$\mathcal{H}_{opt}(q,v,\lambda,\mu,u^{\star})=\min_{u\in\mathcal{U}}\mathcal{H}(q,v,\lambda,\mu, u),$$ then $u^{*}$ may be determined implicitly as a function of $(q(t),v(t),\lambda(t),\mu(t))\in T^{*}\mathcal{D}$ using the previous equation and then we may define the optimal Hamiltonian $\mathcal{H}_{opt}:T^{*}\mathcal{D}\to\mathbb{R}$ by prescribing the control $u$ as $u^{\star}$. 

Given that $u^{\star}$ minimizes $\mathcal{H}$, then $u^{\star}$ is a critical point for $\mathcal{H}$ and may be uniquely determined by \begin{equation}\label{stationary}\frac{\partial\mathcal{H}}{\partial u}(q(t),v(t),\lambda(t),\mu(t),u^{\star}(t))=0,\quad t\in[0,T].\end{equation}

The PMP applied to our particular problem gives the following necessary conditions

\begin{itemize}
\item Stationary condition: from equation \eqref{stationary} $\mu_{A}=-\epsilon u^{A},$
\item State equation: Equations \eqref{control1} and \eqref{control2},
\item Adjoint equations (or costate equations): \begin{align*}
-\dot{\lambda}_i=&\frac{\partial \mathcal{H}}{\partial q^{i}}=(q^{i}-q^{i}_{r})+\lambda_i\frac{\partial\rho_{A}^{i}(q)}{\partial q^{i}}+\mu_A\frac{\partial\dot{v}^{A}}{\partial q^{i}},\\
-\dot{\mu}_A=&\frac{\partial\mathcal{H}}{\partial v^{A}}=(v^{A}-v_r^{A})+\lambda_i\rho_A^{i}(q)+\mu_A\frac{\partial \dot{v}^{A}}{\partial v^{A}},
\end{align*}
\item Constraint induced by terminal cost: $r(\gamma(T), \gamma_{r})=0$,
\item Boundary conditions: $\gamma(0):=(q(0),v(0))\in\mathcal{D}$, 
 $\frac{\partial\Phi}{\partial q^{i}}(\gamma(T))=\lambda_i(T)$,
$\frac{\partial\Phi}{\partial v^{A}}(\gamma(T))=\mu_A(T)$.
\end{itemize}

\subsection{Optimal trajectory tracking for the nonholonomic particle}
Consider the situation of Example \ref{example}. Let $\gamma_{r}$ be the reference trajectory,  $\gamma_{r}=(x_r(t),y_r(t),z_r(t),v_{1,r},v_{2,r})$ which follows the constraint $\dot{x}_r= y_r \dot{z}_r$ at all $t$ and the dynamical equations for the nonholonomic particle. We wish to control the velocity of the nonholonomic particle. We add then control inputs in the fiber coordinates $v^{1}$ and $v^{2}$. Therefore the control dynamical system to study is given by \begin{equation}\label{eqqcontrol}
     \dot{v}^1= u^{1},\quad
     \dot{v}^2=u^{2} -\frac{y}{1+y^2}v^1v^2
   \end{equation} together with  the admissibility conditions $\dot{x}= -y v^2$,
     $\dot{y}= v^1$ and 
     $\dot{z}= v^2.$
     
     The cost function $\mathcal{C}:\mathcal{D}\times\mathcal{U}\to\mathbb{R}$ for the optimal control problem is given by \begin{align*}
\mathcal{C}(q,v, u)= &\frac{1}{2} \left(||x- x_r ||^2 + ||y- y_r ||^2 +||z- z_r ||^2 \right.\\
&\left.+ ||v^1- v^{1}_r||^2 + ||v^2- v^{2}_{r}||^2 + \epsilon ((u^1)^2+(u^2)^2 )\right).
\end{align*} and the terminal cost function is given by \begin{align*}\Phi(x,y,z,v^1,v^2)=
&||x(T)- x_r(T) ||^2 + ||y(T)- y_r(T) ||^2 \\+&||z(T)- z_r(T) ||^2\\
+& ||v^1(T)- v^{1}_r(T)||^2 + ||v^2(T)- v^{2}_{r}(T)||^2\end{align*}with $T\in\mathbb{R}^{+}$ fixed.

The Hamiltonian for the PMP is given as
\begin{align*}
\mathcal{H}(q,v, \lambda, \mu,u)= &\frac{1}{2} \left(||x- x_r ||^2 + ||y- y_r ||^2 +||z- z_r ||^2 \right.\\
+&\left. ||v^1- v^{1}_r||^2 + ||v^2- v^{2}_{r}||^2 + \epsilon (u^1)^2\right.\\
+&\left.\epsilon(u^2)^2 \right)-\lambda_1 y v^2 + \lambda_2 v^1 + \lambda_3 v^2 + \mu_1 u^1 \\&+ \mu_2 \left(u^2-\frac{y}{1+y^2}v^1v^2\right).
\end{align*}

In order for $u(t)$ to be the optimal control we employ the stationary condition. Therefore, $\displaystyle{u_1^{\star} = -\frac{\mu_1}{\epsilon}\; \hbox{ and } \; u_2^{\star} = -\frac{\mu_2}{\epsilon}}$. The final cost is given by $\Phi(\gamma(T))=||\gamma(T)-\gamma_r||^{2}$ which induces the constraint \begin{align*}
r(\gamma(T),\gamma_r)=&||x(T)-x_r ||^2 + ||y(T)- y_r ||^2 +||z(T)- z_r ||^2 \\+& ||v^1(T)- v^{1}_r||^2 + ||v^2(T)- v^{2}_{r}||^2=0.\end{align*}
Finally, the optimal Hamiltonian $\mathcal{H}_{opt}$ is given by 
\begin{align*}
\mathcal{H}(q,v, \lambda)= &\frac{1}{2} \big\{||x- x_r ||^2 + ||y- y_r ||^2 +||z- z_r ||^2 \\
&+ ||v^1- v^{1}_r||^2 + ||v^2- v^{2}_{r}||^2 \big \}-\lambda_1 y v^2 \\&+ \lambda_2 v^1 + \lambda_3 v^2.
\end{align*}

The adjoint equations are  $\dot{\lambda}_1= -(x-x_r)$, $\dot{\lambda}_3 = -(z-z_r)$,  
\begin{align}\label{costateeqn}
\dot{\lambda}_2 &= \lambda_1 v^2 -(y-y_r)+\epsilon v^1v^2\mu_2\left(\frac{y^2-1}{(y^2+1)^{2}}\right),\\ \nonumber
\dot{\mu}_1 &= -\lambda_2 - (v^1- v^{1}_{r})-\mu_2\frac{y}{1+y^2} v^2,\\  \nonumber
\dot{\mu}_2 &= -\lambda_3+ \lambda_1 y - (v_2- v_{2}^{r}) -\mu_2\frac{y}{1+y^2} v_1.
\end{align} The state equations were given in Example \ref{example} in equation \eqref{eqq3} together with the admissibility conditions. Boundary conditions must satisfy the constraints in order for the trajectory to evolve on $\mathcal{D}$, that is $\dot x_0+y_0\,\dot z_0=0$ where $\dot x_0, y_0, \dot z_0$ denotes the boundary conditions for the variables $\dot{x}$, $y$ and $\dot{z}$ respectively.

\subsection{Numerical results}\label{sec5}
We now test with numerical simulations how the proposed methods work.

Denote $F_{\mu}^{\lambda}:[0,T]\times T^{*}\mathcal{D}\to T^{*}\mathcal{D}$, the integral flow given by equations \eqref{costateeqn} on $T^{*}\mathcal{D}$  and $\gamma(0)\in\mathcal{D}$ the initial condition for the state dynamics. The initial guess for the initial condition of the costate variables  is denoted by $ \alpha=F_{\mu}^{\lambda}(0)$. We wish to find the initial condition of the costates for which $F_{\mu}^{\lambda}(T, \gamma(0),  \alpha)=(0_{1\times 5})^T$ . The goal is to find the root of the polynomial \begin{small}\[
F_{\mu}^{\lambda}(\alpha)= \begin{pmatrix}
{\lambda}_1(T, \gamma(0),\alpha)+\omega( x(T, \alpha)- x_r(T)) \\
{\lambda}_2(T, \gamma(0), \alpha) +\omega(  y(T, \alpha)- y_r(T))\\
{\lambda}_3(T, \gamma(0), \alpha) +\omega(  z(T, \alpha)- z_r(T))\\
{\mu}_1(T, \alpha)\\
{\mu}_2(T, \alpha)
\end{pmatrix}
\] \end{small}where $T\in\mathbb{R}^{+}$ is the final time, $\omega\in\mathbb{R}^{+}$ is a weight for the terminal cost and $F_{\mu}^{\lambda}(\tau, \gamma(0), p_0)$ is the flow of the adjoint equations \eqref{costateeqn} starting at $(\gamma(0),p_0)$. The root finder used in both situations was the \textit{fsolve} routine in MATLAB.

\begin{figure}[h!]\label{fig3} \includegraphics[scale=0.17]  {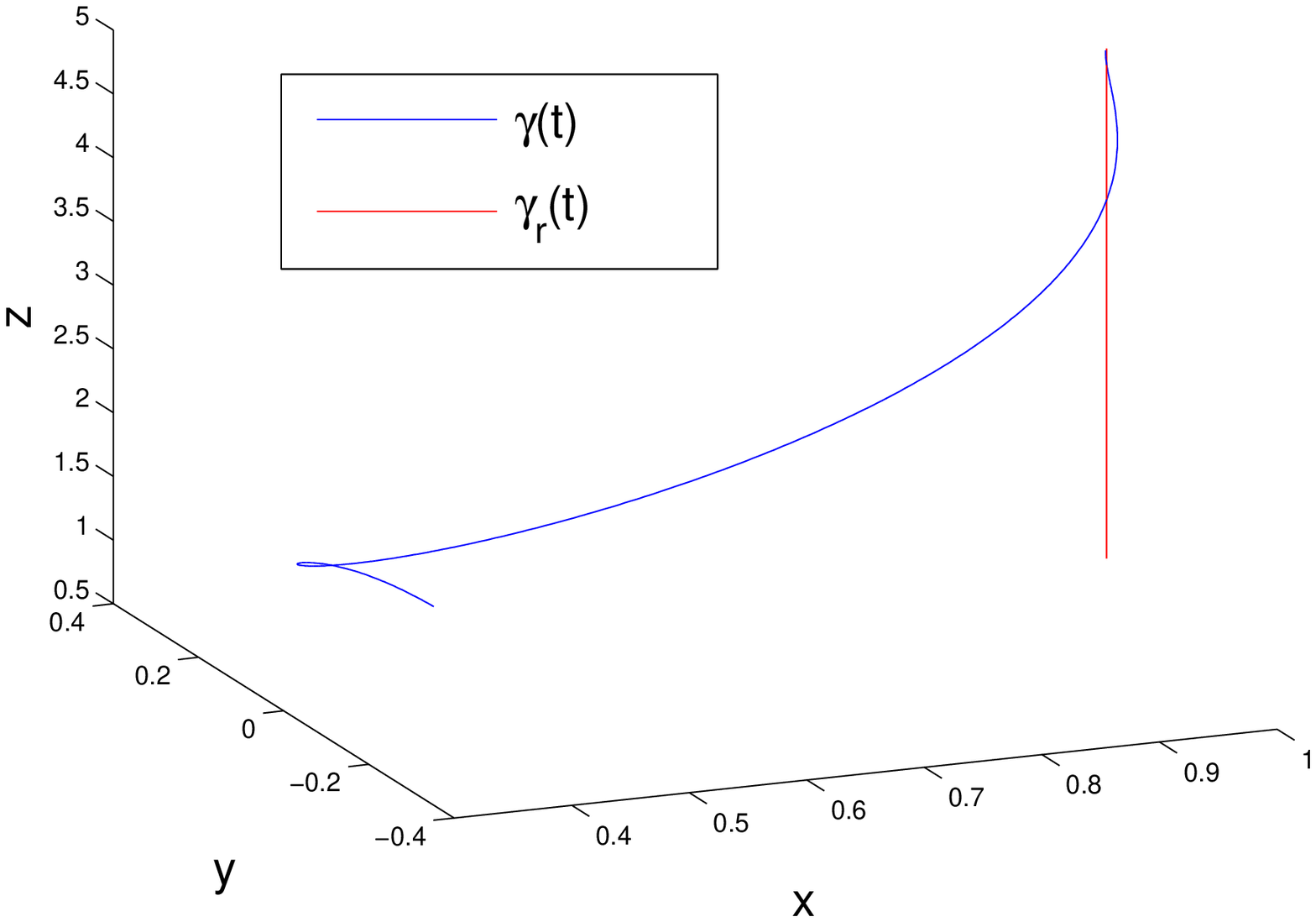}\qquad\quad\qquad\includegraphics[scale=0.17]  {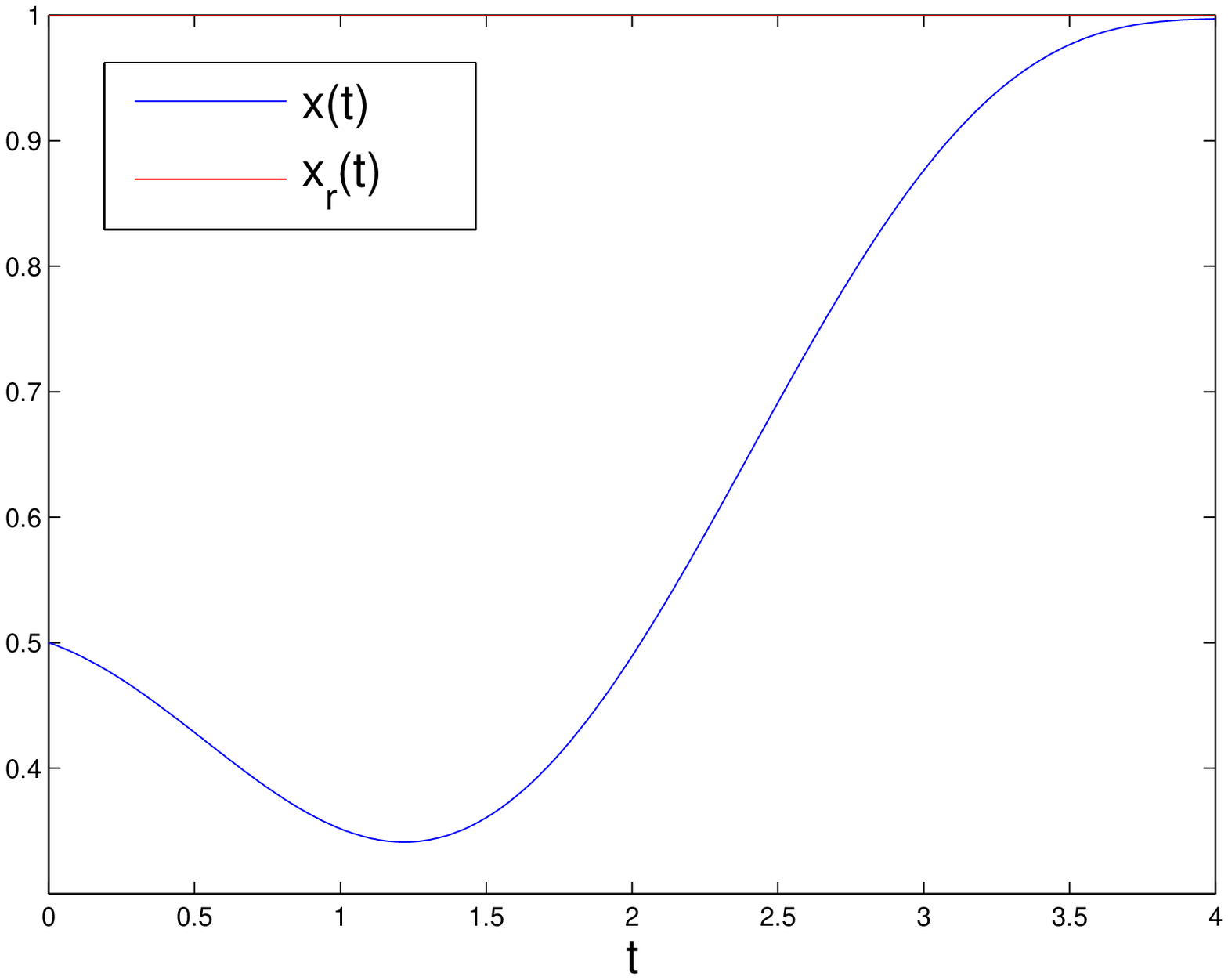}
    \includegraphics[scale=0.17]{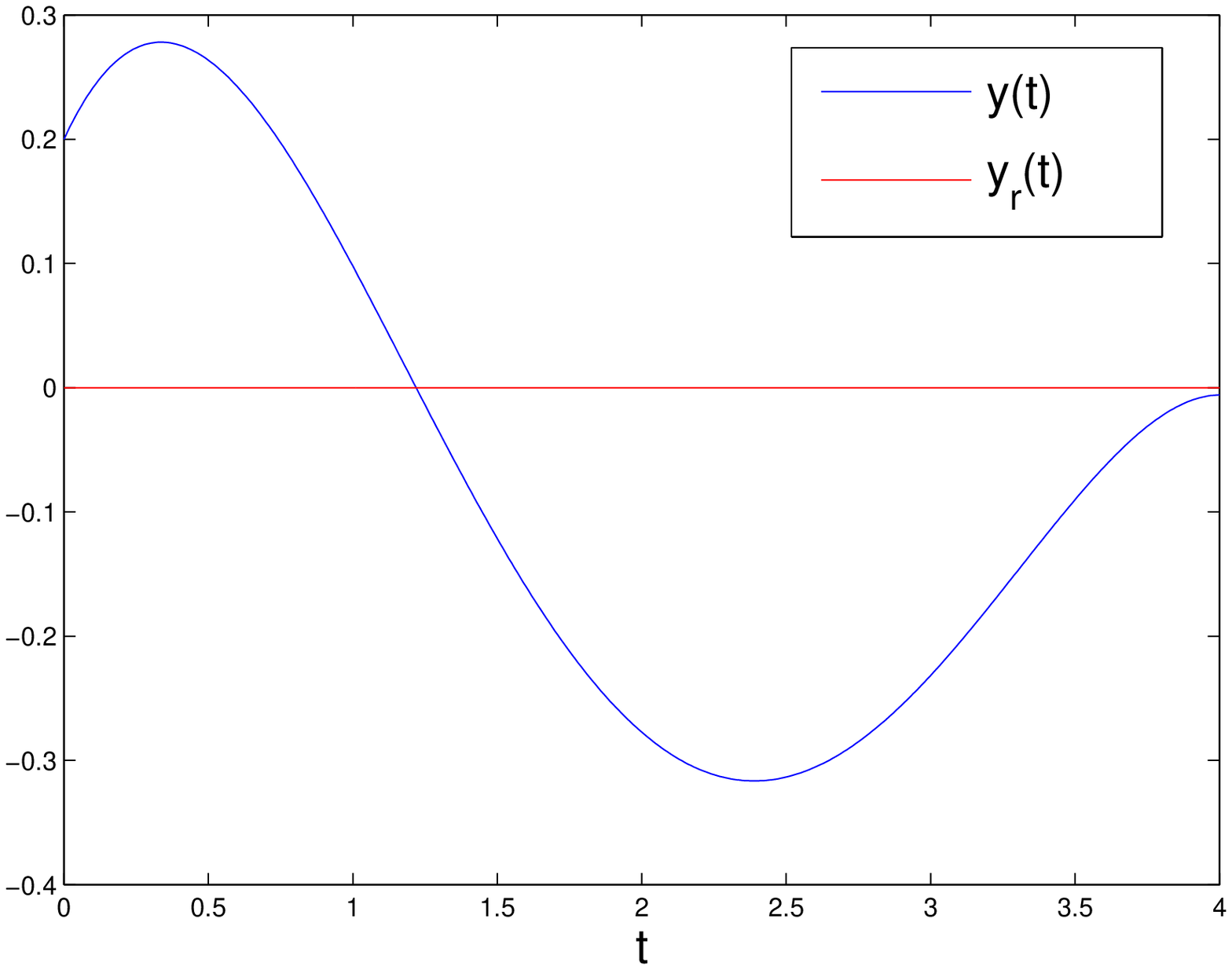}
    \includegraphics[scale=0.17]{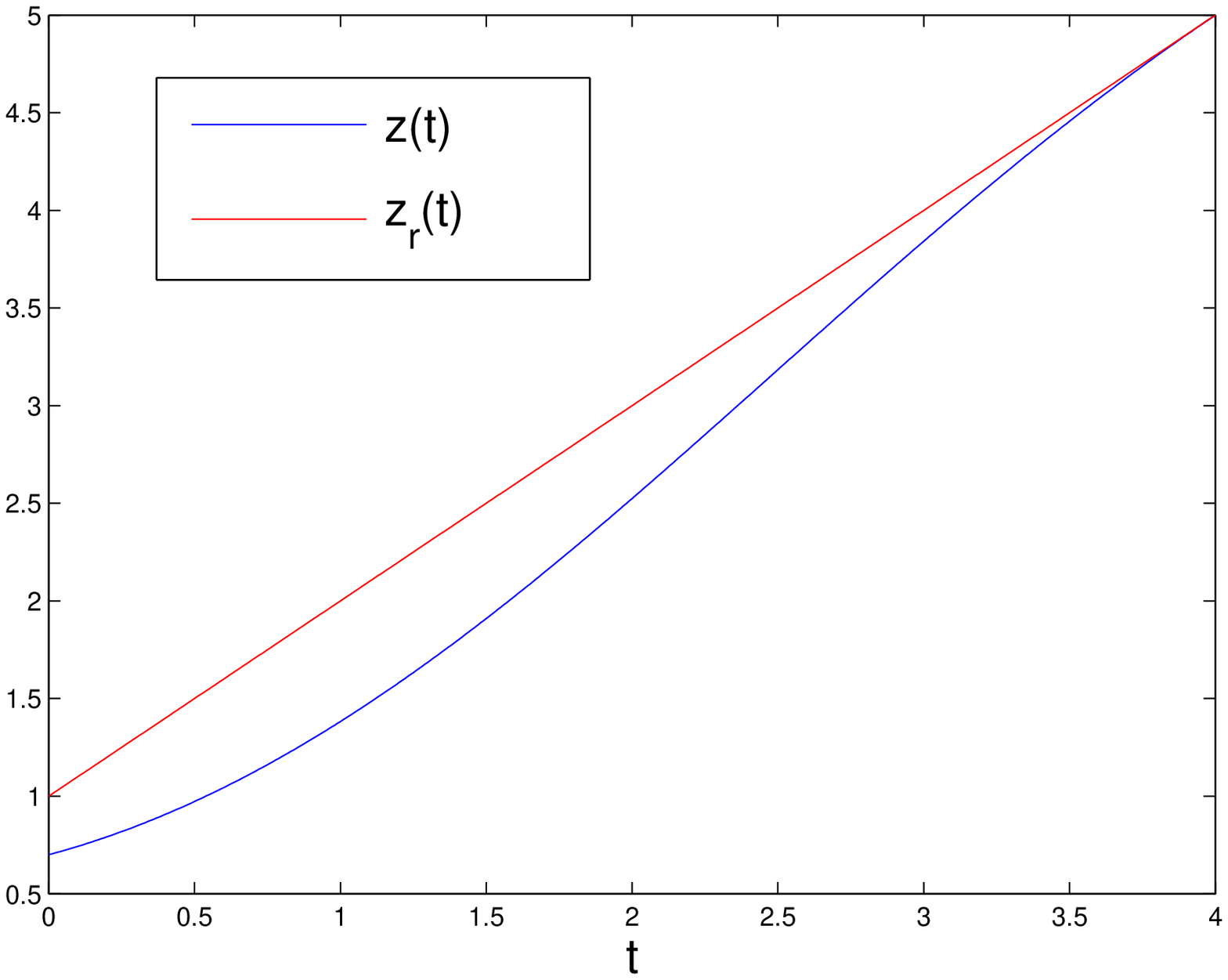}  \includegraphics[scale=0.17]{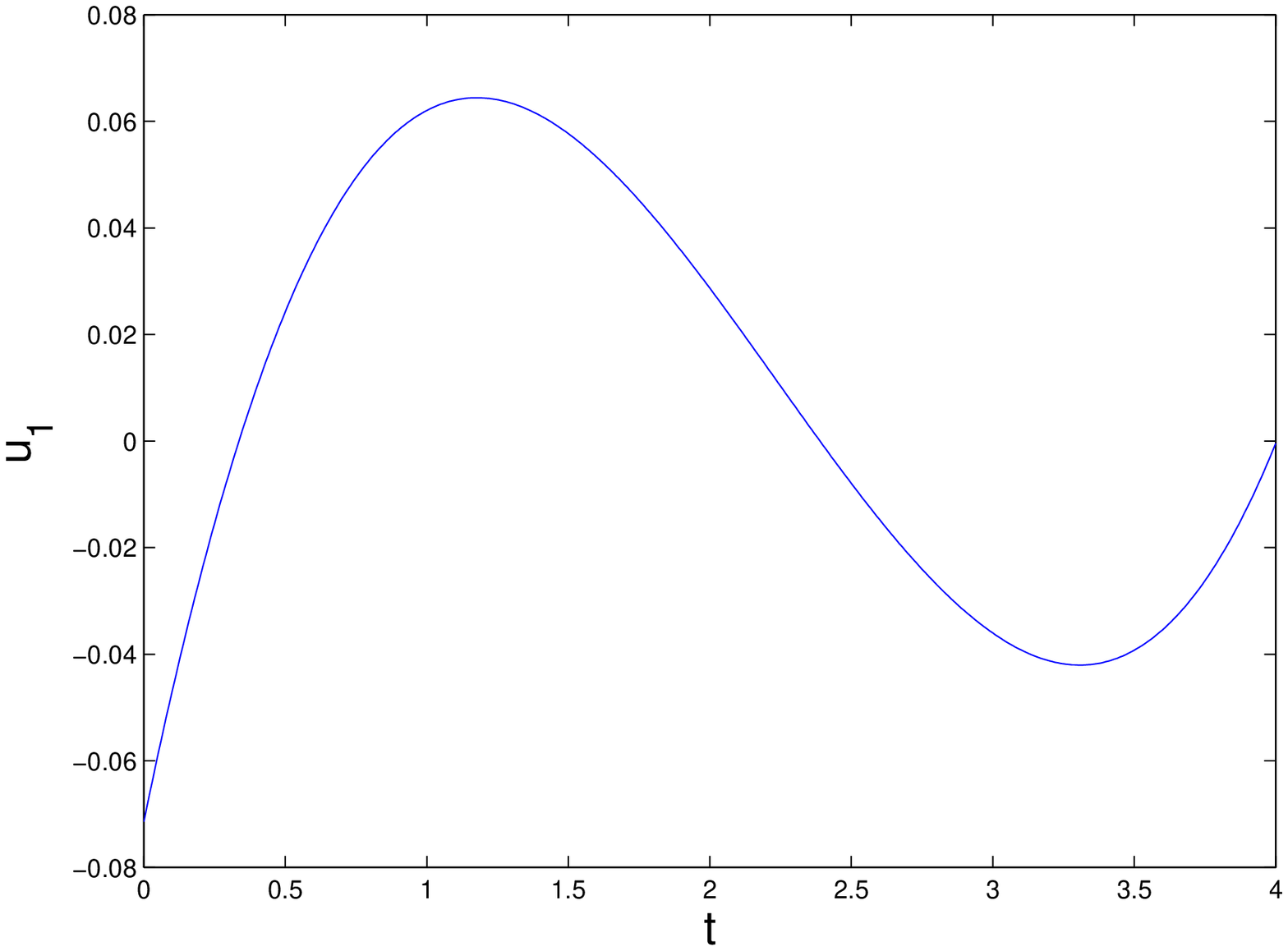}\quad\qquad\qquad  \includegraphics[scale=0.17]{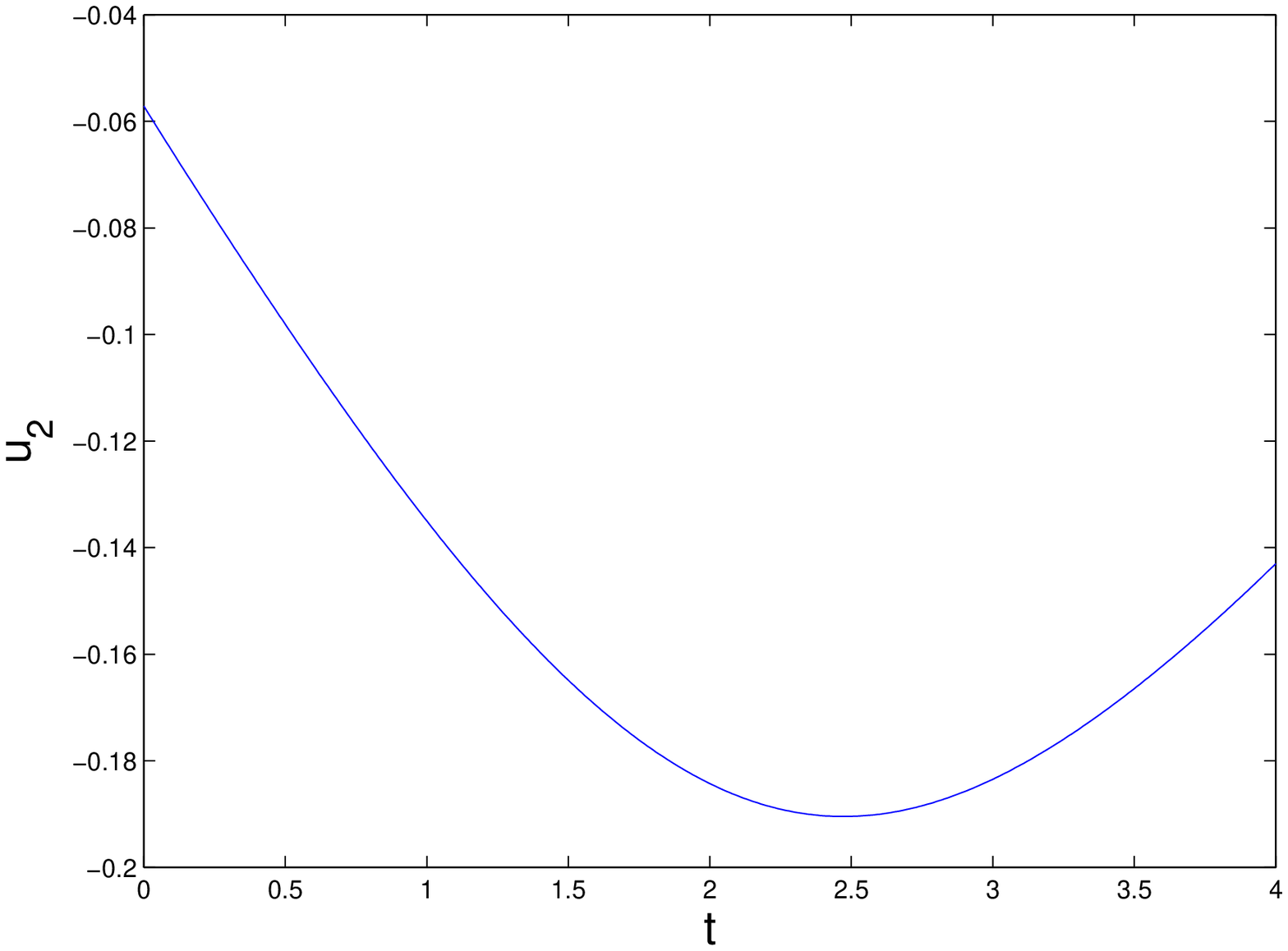}
  \caption{Arbitrary reference trajectory: Trajectories minimizing the cost function $\mathcal{J}$, evolving on $\mathcal{D}$ and tracking the reference trajectory $\gamma_r$ in time $T$ and control inputs}
\end{figure}

For the intial condition $\gamma(0)=$ $\begin{pmatrix}
0.5 & 0.2 & 0.7; & 0.5 & 0.4
\end{pmatrix}$ and reference trajectory $\gamma_r(t)=$ $\left(
1, 0,  t+1,0,1\right) $, $p_0= 0_{1\times 5}$, $T=4$ and $\epsilon=7$ we exhibit the results in Figure \ref{fig3}.

\textbf{Acknowledgments:} The authors wish to thank Prof. Ravi Banavar for fruitful discussions along the development of this work . 

\addtolength{\textheight}{-12cm}   





\end{document}